\documentclass[conference]{IEEEtran}
\IEEEoverridecommandlockouts

\usepackage{cite}
\usepackage{amsmath,amssymb,amsfonts}
\usepackage{algorithmic}
\usepackage{graphicx}
\usepackage{textcomp}
\usepackage{xcolor}
\usepackage{hyperref}  
\usepackage{array} 
\usepackage{titlesec}
\usepackage{afterpage}
\usepackage{lineno}
\usepackage{array} 
\usepackage{adjustbox}
\usepackage{multirow} 
\usepackage{bigstrut} 

\def\BibTeX{{\rm B\kern-.05em{\sc i\kern-.025em b}\kern-.08em
    T\kern-.1667em\lower.7ex\hbox{E}\kern-.125emX}}
\begin{document}

\title{Artificial Intelligence Policy Framework for Institutions}

\author{\IEEEauthorblockN{ William Franz Lamberti}
\IEEEauthorblockA{
\textit{Computational and Data Sciences}\\
\textit{College of Science}\\
\textit{George Mason University }\\
Fairfax, VA, United States \\
wlamber2@gmu.edu\\
williamfranzlamberti@gmail.com
}
}

\maketitle
\IEEEpubidadjcol

\begin{abstract}
Artificial intelligence (AI) has transformed various sectors and institutions, including education and healthcare. Although AI offers immense potential for innovation and problem solving, its integration also raises significant ethical concerns, such as privacy and bias. This paper delves into key considerations for developing AI policies within institutions. We explore the importance of interpretability and explainability in AI elements, as well as the need to mitigate biases and ensure privacy. Additionally, we discuss the environmental impact of AI and the importance of energy-efficient practices.  The culmination of these important components is centralized in a generalized framework to be utilized for institutions developing their AI policy. By addressing these critical factors, institutions can harness the power of AI while safeguarding ethical principles.
\end{abstract}

\begin{IEEEkeywords}
Artificial Intelligence, Generative, Policy, Institutions, Education, National Security
\end{IEEEkeywords}

\section{Introduction}

Artificial Intelligence (AI) has rapidly emerged as a transformative force across various domains, and institutions are no exception. The integration of AI-powered tools, particularly generative AI (GiA), offers users unprecedented opportunities for learning, collaboration, and creativity. However, this rapid evolution also presents significant challenges and ethical considerations. Some have advocated for less laws and policies \cite{thierer_artificial_2017}, while others have advocated for more \cite{greiman_human_2021, hogenhout_framework_2021}.  Furthermore, translating the theory into a practical framework is challenging \cite{zuger_introduction_2024}.  For example, the European Union has passed laws which outlines guidelines for AI use\cite{gille_balancing_2024}.  Regardless, these kinds of guidelines is not universal across all countries and the details of this law within the EU are still being clarified.  This work aims to provide a framework that provides a baseline and practical AI policy framework for institutions.

Governments have considered the use of AI to help speed up the process of administrating decisions on cases.  Brazil utilized an AI to help address more cases more efficiently\cite{nicolas_balancing_2024}.  As an example of the results of this AI, there was a notable drop in cases who were approved for maternity benefits.  However, the AI is a ``black box" and could not clearly state why it had come to the decision it had \cite{nicolas_balancing_2024}.  In some of the states of the United States, some judicial systems have access to COMPASS which assesses the likelihood of a convict committing another crime. This has come under criticism which essentially claims that minorities are mistreated by using COMPASS.  Much work has been committed to assessing the fairness of COMPASS \cite{chatziathanasiou_beware_2022, gursoy_equal_2022}.  It has been suggested that oversight to ensure that protected groups are not continually marginalized is worthwhile \cite{thierer_artificial_2017, calo_artificial_2018, ye_privacy_2024, pasupuleti_cyber_2023}.  

Companies are quickly adopting to the use of AI rapidly.  For example, Google's CEO claimed that about 25\% of new code was written by GiA as of 2024\cite{monge_jim_clyde_more_2024}.  The United States' National Bureau of Economic Research claims that about 18\% of firms have adopted AI in some capacity in 2018 \cite{mcelheran_ai_2023}.  While this adoption was not uniformly distributed across the nation \cite{mcelheran_ai_2023}, this showcases that firms are adopting these technologies across the United States with or without any policies in place. This is not to say that the United States has not been developing policy related to AI, but rather it is not at the same level as the European Union \cite{maslej_artificial_2024}. 

Some have provided an overview of AI policy for governments as it stands.  A major takeaway is that AI driven policies lead to many unexpected consequences \cite{valle-cruz_review_2019}.  Others have noted the general lack of guidance in areas such as education AI policy at a national level\cite{schiff_education_2022}.  However, there has been some review of the AI educational policies for Hong Kong \cite{chan_comprehensive_2023}.  

Public opinion on AI is evolving.  For example, global usage of ChatGPT on a weekly basis is about 36\% \cite{loewen_global_nodate}.  Furthermore, about 33\% and 28\% are using ChatGPT for their jobs and education, respectively\cite{loewen_global_nodate}.  This suggests that ChatGPT is impacting two major components of society given it's release in 2022.  While there are some areas that individuals do not trust AI (such as potential dates of a romantic nature), there are some areas like planning vacations or creating a grocery list where individuals are willing to trust AI results more \cite{loewen_global_nodate}.  Thus, individuals are using AI-based solutions in many different aspects of their lives.  This would then potentially trickle upward and impact our institutions.  

This paper delves into the evolving landscape of AI for institutions (such as education, medicine, and national security), exploring the potential benefits, risks, and essential considerations for institutional policy development. We will examine how AI can enhance learning and work experiences, foster innovation, and address disparities. At the same time, we will discuss the ethical implications of AI, including privacy concerns, bias, and the potential for misuse.

By understanding the complexities of AI for institutions, we can develop a comprehensive policy framework that guide its responsible and effective implementation. These policies should strike a balance between harnessing the potential benefits of AI while mitigating its risks, ensuring a future where AI serves as a powerful tool for institutional advancement.

\section{Context for AI}

It is important to define some terminology and provide definitions.  These definitions will help to characterize and better understand what items we are and are not mentioning.  The three major items are: computers, AI, and GiA.  There does not appear to be a unified consensus on these definitions\cite{thierer_artificial_2017}.  We recognize that some might have differing definitions, but we will utilize the following definitions and descriptions for the purposes of this article. 

\subsection{Computers}

A \textbf{computer} is an electronic device that processes instructions.  Computers serve as a fundamental tool for AI development. Computers are employed across a broad spectrum of applications, ranging from basic calculations and tasks (such as counting time) to intricate calculations (such as calculating a derivative). Some examples of computers using this definition includes items like microwaves, smartphones, and laptops.  Each have different specific uses and capabilities, but all are computers for our discussion.  

\subsection{Artificial Intelligence}
For the purposes of this manuscript, \textbf{AI} is defined as an operation that performed by a computer that could be performed by a human.  In other words, AI refers to a computer's ability to emulate human cognitive functions, such as learning and problem-solving.  This differs from some definitions of AI, which seem to be synonymous with deep learning or a neural network-based approach \cite{hogenhout_framework_2021}.  

While computers are integral for AI, they are not synonymous with AI.  It's important to note that all computers are equipped with some kind of AI capabilities. With the recent significant advancements with various techniques(such as deep learning), AI has expanded its horizons to encompass relatively complex endeavors, including generating realistic images\cite{wang_diffusiondb_2023, xu_imagereward_2023} and translating languages \cite{luccioni_power_2024}.

\subsection{Generative AI}

\textbf{GiA}, a specialized branch of AI, focuses on the creation of content with a spatial component. This includes images, text, or music. Generative AI models are trained on extensive and large datasets of existing content, enabling them to generate new, original content based on user provided prompts \cite{luccioni_power_2024}. For instance, a generative AI model trained on a dataset of paintings can subsequently generate new paintings in a similar artistic style.

\section{Key Considerations for AI Policy}

There are a number of different components to consider when it comes to using a framework for AI policy on a specific application.  However, they all regard an element or multiple elements.  In this context, an \textbf{element} is a essential part of a product or solution that utilizes AI.  For example, an element could be the type of metric used to describe a characteristic.  An element could also be more intricate such as an algorithm applied to a specific problem within an institution.  Regardless, all AI elements have the same key qualities to consider when weighing the costs.

\subsection{Explainable and Interpretable AI}

There are a number of key components required for proper decision making when choosing how to use AI for institutions.  A powerful method of determining the appropriate use of a given AI method is the amount of explainability and interpretability it has \cite{barredo_arrieta_explainable_2020, lamberti_william_franz_overview_2022, hogenhout_framework_2021}. \textbf{Explainability} is the property of an AI that ``allows its mechanisms to be explicitly described, understood, and studied"\cite{lamberti_william_franz_overview_2022}. \textbf{Interpretability} is the property of an AI ``to have concrete physical meaning"\cite{lamberti_william_franz_overview_2022}. These pillars of interpretability and explainability help to establish trust between experts and users. 

Trust is a critical component of successful AI implementation within institutions \cite{gille_balancing_2024, barredo_arrieta_explainable_2020}. It is essential to build trust between AI systems, AI experts, and users of AI \cite{barredo_arrieta_explainable_2020}. To foster trust, AI systems must be interpretable and explainable. Trust involves providing clear information about the AI system's capabilities, limitations, and decision-making processes. Explainability and interpretability enables users and experts of AI to the system's outputs, thereby increasing confidence and trust. 

An analogy: during the development and expansion of electricity during the late 1800's and early 1900's Edison's\cite{edison_thomas_1884} and Tesla's\cite{tesla_electro-magnetic_1888} two competing technologies (DC and AC, respectively), were competing in scientific thought and market adoption in what is sometimes called ``War of the Electric Currents''\cite{jonnes_empires_2003}.  Many of these discussions and experiments were based on the fact that there wasn't a clear level of trust in DC and AC in the expert community.  With over 100 years of studying DC and AC technologies, we now have a higher level of trust because the technologies can be explained and interpreted at a higher level than in the late 1800s.  

Many AI elements are still in their embryonic stages; similar to our understanding of electricity was in the late 1800s.  To help get to a higher level of trust for some AI elements, we need a stronger level of interpretability and explainability of said AI elements.  While it is outside the scope of this manuscript to evaluate many AIs, a broad overview has been accomplished \cite{lamberti_william_franz_overview_2022}.  We will focus on some examples of algorithms which are the fundamental backbone of many AI solutions to help explain this work briefly herein. 

Neural Networks (NNs) are powerful tools for various tasks such as image classification \cite{fukushima_neocognitron:_1980}. However, they often suffer from two significant drawbacks: low interpretability and low explainability.  NNs typically involve a large number of parameters, making it difficult to understand how they arrive at a specific prediction \cite{lamberti_william_franz_overview_2022}. The complex interactions between these parameters often result in a "black box" model, where the internal workings are unclear.  NNs can approximate a wide range of functions, but the exact function being learned by a specific NN is often unknown. This lack of transparency makes it challenging to explain the rationale behind a particular prediction, hindering trust and understanding.  These limitations are particularly pronounced in deep NNs, including Artificial Neural Networks (ANNs) \cite{rosenblatt_perceptron_1958} and Recurrent Neural Networks (RNNs) \cite{rosenblatt_principles_1962}, which often have millions of parameters \cite{sahlol_efficient_2020}. While these models achieve impressive performance, their complexity makes it difficult to decipher the underlying mechanisms that drive their decisions.

Ordinary Least Squares (OLS) or regression stands in stark contrast to NNs in terms of interpretability and explainability. Unlike the "black box" nature of NNs, OLS offers a high degree of transparency in understanding how it arrives at predictions \cite{barredo_arrieta_explainable_2020}.  OLS has a high level of explainablity since it offers a clear description of the overall relationship between variables \cite{lamberti_william_franz_overview_2022}.  The core equation (i.e., $Y = \beta X$) explicitly depicts the linear relationship between the dependent variable ($Y$) and the independent variables ($X$) \cite{mendenhall_second_2011}. This allows for a clear understanding of how changes in the independent variables collectively influence the dependent variable.  Additionally, the $R^2$ value provides a measure of how well the model explains the overall variation in the data \cite{mendenhall_second_2011}.  OLS has a high level of interpretability since each coefficient associated with an explanatory variable ($X$) has a clear and direct meaning \cite{lamberti_william_franz_overview_2022}.  For example, a coefficient of 10 for a given input variable translates to: holding all other variables constant, a one-unit increase in the input variable is expected to lead to a 10-unit increase in the output. This straightforward mapping from coefficient to effect size allows for easy interpretation of the model's results \cite{lamberti_william_franz_overview_2022}.

Explainability and interpertability provide the means by which we can evaluate AI elements.  By using these concepts, we can better understand which AI solutions are better suited for different scenarios such as universities or medical clinics. 

\subsection{Privacy}

The integration of AI in institutions presents a complex interplay between technological advancement and ethical considerations, particularly concerning privacy and data security \cite{pasupuleti_cyber_2023, chik_generative_2024, ye_privacy_2024}. As AI systems increasingly rely on large datasets to learn and make informed decisions, safeguarding sensitive information becomes paramount.

For example, the use of personally identifiable information (PII) as input for AI systems poses significant privacy risks. PII (such as social security numbers, addresses, or biometric data) can be highly sensitive. Compromising PII for individuals can lead to identity theft\cite{pasupuleti_cyber_2023}, financial loss \cite{ye_privacy_2024}, or reputational damage \cite{chik_generative_2024}. To mitigate these risks, institutions should prioritize the use of anonymized or pseudonymized data, where personal identifiers are removed or replaced with unique codes.  For example, academic institutions should not include the names or academic IDs values as inputs to publically available GiA models to ensure that a student's private information is not compromised.  

\subsection{Bias and Fairness}

In addition to data privacy, AI systems must be designed and implemented in a way that avoids perpetuating biases and discrimination \cite{chik_generative_2024}. AI algorithms are trained on data, and if that data is biased, the AI system will also be biased. This can lead to unfair and discriminatory outcomes, particularly in areas such as hiring\cite{chik_generative_2024}, lending \cite{chik_generative_2024}, and criminal justice \cite{chatziathanasiou_beware_2022, gursoy_equal_2022}. To address this issue, it is crucial to use diverse and representative datasets, and to regularly monitor and audit AI systems for bias.  

To ensure that fair results are achieved, institutions need to ensure that the model does not impact different groups differently (especially protected groups).  For example, academic institutions should not use features that correspond or could correlate with (i.e., skin color) with race for determining academic honor code violations.  

\subsection{Sustainability}

The integration of AI into various sectors raises important sustainability concerns. As AI models become more complex and sophisticated, the computational resources required to train and run them increase exponentially, particularly for GiA solutions\cite{luccioni_power_2024}. This surge in energy consumption can contribute to increased demand for resources. 

To mitigate these energy impacts, it is crucial to adopt energy-efficient algorithms, optimize hardware utilization, and promote responsible AI practices. Researchers and developers should prioritize the development of algorithms that require less computational power and energy for non-essential tasks.  

\section{Framework for AI Policy Development}

\begin{figure*}
    \centering
    \includegraphics[width=\linewidth]{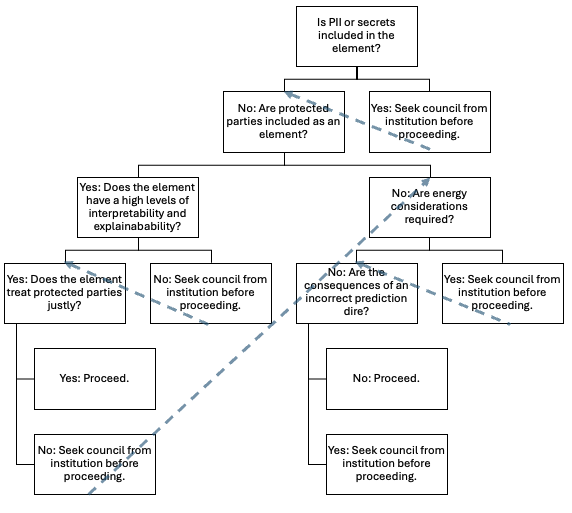}
    \caption{General AI policy for institutions.}
    \label{fig: ai_flow}
\end{figure*}

To ensure the ethical and responsible use of AI within an institution, a decision-making framework can be employed to assess the potential implications of various AI applications. Here is a breakdown of the decision points outlined in the flow chart in Figure \ref{fig: ai_flow}:

1. PII and Secret Inclusion:  The first step involves determining whether the elemenet involves personally identifiable information (PII) or secrets. If PII or a secret is included, careful consideration must be given to data privacy and security. Robust measures should be implemented to protect sensitive information, such as encryption, access controls, anonymization techniques, or removal.

2. Protected Parties: If the element does not involve PII, the next step is to assess whether it affects protected parties, such as individuals with disabilities or minority groups. If protected parties are involved, it is crucial to ensure that the element is fair, unbiased, and does not perpetuate discrimination.

3. Interpretability and Explainability:  For elements that involve protected parties, high levels of interpretability and explainability are essential. This means that the decision-making process of the element should be transparent and understandable. Interpretable and explainable elements can help identify and mitigate biases, ensuring fair and equitable outcomes.

4. Energy Considerations: If the AI application is computationally intensive, such as large-scale generative AI models, it is important to consider the energy implications. High energy consumption can contribute to environmental impact and increased costs for institutions. 

5. Consequences of Incorrect Predictions: Finally, the potential consequences of incorrect predictions should be evaluated. If the consequences are severe, such as in healthcare or national security applications, it is crucial to prioritize safety and reliability. This may involve additional testing, validation, and human oversight to minimize the risk of errors.

To use the decision flow chart in Figure \ref{fig: ai_flow}, start at the top and follow the "Yes" or "No" branches according to the specific characteristics of the element. First, determine if the element involves PII or secret information. If yes, additional safeguards must be implemented to protect sensitive information.  Once those safegaurds are obtained, one may act as if the opposite answer was given (i.e., No).  In other words, once council has been given and approval received, follow the dotted line to the next step in the decision flow chart.  

The next question is to consider whether the application affects protected parties. If it does, the next question is related to the level of explainability and interpretability of the element.  If it does not, the next question considers the energy requirements of the element.  This process continues until termination at a node with the words "Proceed".  

\begin{figure*}
    \centering
    \includegraphics[width=\linewidth]{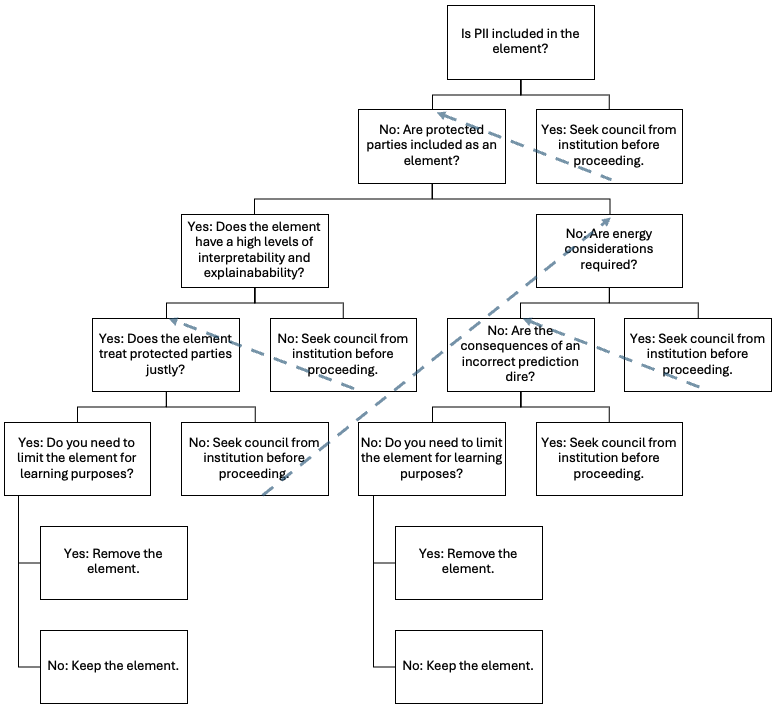}
    \caption{Altered AI policy for classroom settings.}
    \label{fig: ai_class}
\end{figure*}

The academic institution has a unique situation.  In some cases (such as the classroom), it may want to limit the use of AI elements to help students learn how to do certain tasks manually.  For example, an instructor may limit the use of a calculator in a mathematical heavy class so that the students learn how to do the arithmetic or calculus by hand.  Another example is limiting the use of GiA to write papers.  (An instructor may achieve this my having the students write essays in-person in a blue book.)  Figure \ref{fig: ai_class} showcases this unique situation with additional steps to consider. 

The following case studies showcase how one might use this decision flow chart given different specific applications.  While the details of each are hypothetical, they are intended to be believable.  The case studies main goal is to be used to navigate Figure \ref{fig: ai_flow}.  Thus, only those details which help inform navigating the decision chart are emphasized and described so that a wide audience understands the main points in a succinct manner. 

\subsection{Case Study: Video Game Graphics}

A video game studio wants to ensure that water looks and behaves more realistically in their upcoming video game.  They are interested in using a deep learning NN to achieve this, but are unsure if it is appropriate to use in this context.  They consult the Figure \ref{fig: ai_flow} as a guide.  

Their application does not have any PII or secrets, as it is a model to make more realistic water and the model was publically available on an open source repository.  Their application does not include any protected parties, as it is a model for realistic water. The model is also fairly efficient, so there are no energy concerns.  There are also no dire consequences of a bad outcome.  In other words, no harm would come to person, organization, or group if the water did not look or behave as realistically as desired. One could argue that if the result was so glitchy and poorly implemented that the game was unplayable, the game studio might not have a lot of sales and therefore individuals might lose their job (which would be a dire consequence).  However, we are assuming that the NN is working as intended and at an acceptable level.  

\subsection{Case Study: Academic Honor Code Violation}

Assume that at a given university, the honor code violation staff is inundated with cases to get through.  It is suggested that they utilize an AI model to perform the classifications of each case to get through them more efficiently.  This model's output predicts a guilty or not guilty classification or label for each case.  However, each case has the student's ID number with the case.  After the staff speak with the appropriate councils, it is determined that the ID number can be utilized without any alterations as it is used only to communicate with the associated individual associated with the case and all communications are kept internally.  One of the AI model's input variables is the race of the associated student.  The input variable is highly interpretable and explainable as it is the self-identified race of the associated student.  However, when comparing the model to the data used to train the model, it is found that those of a certain races are overclassified as guilty.  In other words, the AI model does not treat certain protected groups justly.  After reporting this finding to the honor code violation staff supervisors, the supervisors recommend not using the AI model.  The cost of a slower process but ensuring that protected groups are treated justly outweighs any gains of processing cases more efficiently.  

\subsection{Case Study: Diagnosing Blood Cancers at a Clinic}

A medical clinic specializing in diagnosing leukemia at a clinic wants to use an AI model to provide classifications if a patient has leukemia.  The model predicts ``leukemia" or ``not leukemia".  The model does not take into account any PII, but it does include many personal health information for many patients.  The model will not be published outside of the clinic, all proper authorizations for using the patient data has been obtained, and the data and model has the necessary security measures in place to prevent the model and patient data from shared outside of the clinic.  While race is not included as feature input, known past family members with particular blood disorder called shovel cell disease is included because that disorder has a large effect on how the image data is portrayed.  However, shovel cell disease is strongly correlated with particular protected groups of a particular race.  Every aspect of the AI model is highly interpretable and explainable from the algorithm, input features, and evaluation metrics.  While those with shovel cell disease are not classified as well as those without shovel cell disease, the AI model still greatly improves the quality of life of all patients used for future predictions.  Thus, the model is adopted within the clinic and the medical staff is informed of the proper use of using the AI model.

\subsection{Case Study: Diagnosing Blood Cancers in an Academic Journal}

A professor at a Statistics department strongly desires to help push the boundaries of scientific knowledge regarding diagnosing leukemia.  Her research involved using an AI model to provide classifications if a patient has leukemia.  The model predicts ``leukemia" or ``not leukemia" and uses a deep NN.  The model does not take into account any PII, but it does include many personal health information for many patients.  The model will be published in an academic journal, all proper authorizations for using the patient data has been obtained, and the data she used to build the model comes from a well known open source repository.  Only the image data is used to build the model, so information regarding protected groups is not included as inputs.  Since a deep NN is used, the model is not interpretable nor explainable.  Furthermore, the model severely misclassified patients over the age of 75.  However, she notes in her paper that this model should not be adopted at a clinical stage due to this issue, but it is listed as an aspect to be further improved in future iterations of the model.  Energy considerations are not a concern as she is trying to perform develop a model that will optimize classifying patients as having leukemia or not.  If the best model requires large amounts of energy to build, that is an acceptable trade-off.  Thus, the model is published in a top journal as it helps to push the boundaries of our scientific understanding and bettering of our world.

\subsection{Case Study: Sending an Email using GiA at a Military Institution}

An administrator at a military institution often interacts with highly sensitive information in her email.  She is often sees content that is considered ``Top Secret".  She recently received a rude email from her supervisor about finalizing a report.  In an effort to respond quickly and promptly, she uses GiA to help draft a polite response.  The prompt used by the GiA provides some context, but does not include any names nor any top secret information.  The content of the email is purely administrative and protected parties are not included.  Energy considerations are not considered since speed and efficiency is required.  Lastly, the consequences of the output from the GiA is not dire as the administrator is using this to draft an initial email.  Thus, she uses the GiA to draft the initial email quickly and promptly.  The draft was used almost as is, with only some slight modifications and without citation.  

\subsection{Case Study: Calculator Use in a Calculus Class}

An instructor at a university teaches a Calculus class.  The goal of the class and assignments is to be able to do calculations by hand.  She is trying to determine if calculator use should be permitted in her class or not. (Calculators are an AI element since they replace humans in performing arithmetic by using a specialized computer.)  Calculators do not take any PII, secrets, or protected party information as inputs.  Energy use is not a concern in this case, and the consequences are not dire if the calculator provides an incorrect value (i.e., human error in typing the commands into the calculator).  However, the goal of the course is to be able to do Calculus by hand.  Thus, while the professor allows for students to use calculators of any kind during homework assignments, she limits their use on exams so that the learning outcomes of the course are satisfied.

\subsection{Case Study: GiA Use in a Graduate Class}

An instructor at a university teaches a graduate history class where longer-form typed papers are typical.  The goal of the class and assignments is to perform analytical analyses.  As always, he will evaluate what is provided in the submitted paper (not prior drafts or what the student intended to write).  Prior to the advent of GiA, he always allowed and encouraged students to have their worked edited by their peers or writing tutors.  He is now trying to determine if GiA use should be permitted in his class or not. We will assume that the GiA does not take any PII or secretes as inputs.  The protected parties could be the subject of a student's paper (i.e., the treatment of indentured Japanese in consecration camps in the United States of America during World War II).  However, he is expecting students to evaluate and analyze their papers before submission.  Thus, if students submit factually incorrect information as part of their paper, he will deduct points from the paper due to the error.  The GiA does not have high levels of interpretability and explainability.  However, he expects students to use the output of GiA as an initial draft for their all or sections of their submitted papers.  Again, he expects the students to be responsible for their submissions.  It is unclear if the element treats protected parties justly (as the GiA has been known to provide the rare output that mischaracterize protected parties with fictitious sources).  The instructor expects that students will take responsibility for their submitted statements.  By providing a paper with fictitious sources, the student is being careless and using a tool incorrectly.  The instructor will deduct points from the paper for using inaccurate information and falsified sources.  Energy considerations are not a consideration in this case.  Consequences are not dire for obtaining an incorrect output (i.e., no loss of life) since the student is fully capable of altering the output before submission.  The professor does not need to limit the element for learning purposes as the professor expects his students to be able to write intelligent statements from graduate students.  Thus, the professors determines that the GiA does not take away from the goals of the course, but helps the students to be more efficient.

\section{Conclusion}   

The integration of AI into institutions presents both immense opportunities and significant challenges. By adopting a comprehensive policy framework that considers interpretability, explainability, fairness, privacy, and sustainability, institutions can harness the power of AI while mitigating its risks.  The decision flow chart presented in Figure \ref{fig: ai_flow} provides a valuable tool for navigating the complexities of AI implementation, ensuring that elements utilizing AI are used ethically and responsibly. 

The case studies showcase how the decision flow chart guides institutions through various scenarios. From video game graphics with minimal risk to honor code violation detection with privacy concerns and racial bias, to medical cancer diagnosis with ethical considerations, the flow chart helps institutions make informed decisions. Future work could entail utilizing complexity science approaches (such as agent-based modeling) to simulate these public policy case studies \cite{furtado_policy_2019}. 

For academic or educational institutions, an additional consideration is warranted.  There may be situations that warrant the use of no computers and/or AI.  For example, in elementary school institutions, no calculators may be used in math classes so that students can learn fundamental arithmetic.  Another example is where students are recording themselves playing instruments.  In this context, while a computer may be needed, generative AI to create the desired song defeats the purpose of learning how to do the task.  In essence, there may be situations where AI should be prohibited in classroom settings. 

By promoting transparency, accountability, and responsible development, institutions can leverage AI as a force for positive change in various sectors. Through careful planning and ethical considerations, AI can empower institutions to address complex challenges, enhance learning experiences, and advance human progress. 

\section{Acknowledgments}

The following tools were used to write this manuscript: Overleaf, Writefull (which is built into Overleaf as of 2024), Microsoft Word, Notability on an iPad, 1st generation Apple Pencil, Macbook Pro personal and work computers, Ubuntu desktop homebuilt desktop computer, Gemini, Zotero, Preview on Mac, Safari, and Firefox.  Tools not mentioned were unintentionally omitted.  

\clearpage

\bibliographystyle{IEEEtran}
\bibliography{main.bib}

\end{document}